\documentclass[floatfix,aps,showpacs,twocolumn,nofootinbib,preprintnumbers]{revtex4}
\usepackage{graphicx}
\usepackage{epsfig}
\usepackage{bm}
\usepackage{longtable}

\def\lsim{~\rlap{$<$}{\lower 1.0ex\hbox{$\sim$}}\;}
\def\gsim{~\rlap{$>$}{\lower 1.0ex\hbox{$\sim$}}\;}

\newcommand{\gev}{\mbox{ GeV}}

\begin{document}


\title{A new signature for color octet pseudoscalars at the LHC}

\author{Alfonso R.~Zerwekh}
\email{alfonsozerwekh@uach.cl}
\affiliation{Instituto de F{\'{\i}}sica, Facultad de Ciencias,
Universidad Austral de Chile, Casilla 567, Valdivia, Chile}

\author{Claudio O.~Dib}
\email{claudio.dib@usm.cl}
\affiliation{Department of Physics, Universidad T\'ecnica Federico
  Santa Mar\'\i a, Valpara\'\i so, Chile}

\author{Rogerio Rosenfeld}
\email{rosenfel@ift.unesp.br}
\affiliation{Instituto de F\'{\i}sica Te\'orica - UNESP, Rua
Pamplona, 145, 01405-900, S\~{a}o Paulo, SP, Brazil}

\date{\today}

\begin{abstract}
Color octet (pseudo)scalars, if they exist, will be copiously
produced at the CERN Large Hadron Collider (LHC). However, their
detection can become a very challenging task. In particular, if
their decay into a pair of top quarks is kinematically forbidden,
the main decay channel would be into two jets, with a very large
background. In this Brief Report we explore the possibility of
using anomaly-induced decays of the color octet pseudoscalars into
gauge bosons to find them at the LHC.
\end{abstract}


\maketitle

\section{Introduction}

In spite of its great experimental successes \cite{EWWG}, the
standard model (SM) of the electroweak interactions is still
widely regarded as an incomplete theory. The reasons are manifold,
including the existence of non-baryonic dark matter and non-zero
neutrino masses in addition to the theoretical problems of
triviality and naturalness related to the scalar Higgs sector
responsible for the mechanism of electroweak symmetry breaking.

There are many extensions of the SM that require the existence of
color octet scalar particles, such as the extra component of the
gluon field in models with extra dimensions \cite{extragluon} or
supersymmetric models with an adjoint chiral supermultiplet
\cite{susyadjoint}. The existence of color octet scalars have also
been used to explain the accelerated expansion of the universe
\cite{Dejan}. Extended scalar sectors of the SM with color octet
scalars that respect the principle of minimal flavor violation
were also recently considered \cite{ManoharWise,GreshamWise}.
Color octet scalars may as well have important effects in the
Higgs boson production via gluon fusion \cite{Bonciani}.

Early studies on the existence of color octet scalars were done in
the context of one-family technicolor models \cite{1family}. Both
electroweak triplets ($P_8^{\pm,0}$) and singlets ($P_8^{0
\prime}$, sometimes denoted also as technieta $\eta_{T8}$, a notation which we
will adopt in this paper) are
present in the spectrum of the pseudo-Nambu-Goldstone boson (PNGB)
arising from the global $SU(8)_L \times SU(8)_R \rightarrow
SU(8)_V$ spontaneous symmetry breaking \cite{review}. The masses
of the color octet PNGB arise mainly from QCD contributions and
are expected to be of the order of $300$ GeV \cite{masses}.

The cross section for pair production of the color octet scalars
is dominated by gluon fusion, which in turn is determined by gauge
invariance \cite{ZR} and consequently, except for the gluon parton
densities, is mostly model independent, being fixed by the masses
of the particles. There could also be a model dependent
enhancement due to the coupling of the scalars to color-octet
vector resonances such as a technirho $\rho_{T8}$ \cite{EHLQ},
which in turn couples to quarks and gluons. However, as shown by
two of the present authors \cite{ZR}, a proper gauge invariant
treatment of the $\rho_{T8}$ results in a vanishing
$\rho_{T8}$-$g$-$g$ coupling \cite{CasalbuoniEBESS}.  Hence, only
the quark initial state can cause this enhancement.

While models of a strongly interacting sector responsible for
electroweak symmetry breaking fell in disfavor in the mid nineties
due to tight bounds from electroweak precision measurements, they
have experienced a recent resurrection due to the correspondence
with weakly interacting models in extra dimensions
\cite{CompHiggs}. In this case, the $\rho_{T8}$ could be
interpreted as the Kaluza-Klein excitation of the gluon.

Pair production of color octet scalars at the LHC have been
recently analyzed \cite{ManoharWise,GreshamWise,Dobrescu,Gerbush}.
Their detection was discussed using the $b \bar{b} b \bar{b}$, $b
\bar{b} t \bar{t}$ and $t \bar{t} t \bar{t}$ channels
\cite{Dobrescu,Gerbush,Lillie:2007hd} for a scalar mass of the
order of a TeV (a recent analysis using a generic four-jet channel
at the Tevatron is discussed in \cite{Kilic}). However, the low
mass values suggested by technicolor models, typically below the
$t \bar{t}$ threshold, can be more challenging to observe. We find
it timely to extend the work done previously in \cite{ZR}, where
rarer decay modes of the color octet (pseudo)scalar are used,
namely decays induced by the chiral anomaly into massless gauge
bosons (photons and gluons), to investigate the possibility of
detecting these new states at the LHC.

In section II we describe the details of our model, in particular
the description of $\eta_{T8}$ pair production and decay. In
section III we present the details and results of our simulations
of $\eta_{T8}$ pairs, produced in $p-p$ collisions at LHC energies
and detected as $\gamma+ 3$jet events; we include a study of the
background and the cuts required for a statistically significant
signal. Finally in section IV we draw our conclusions.

\section{Model}

When introducing massive vector bosons in a theory that contain
gauge fields, care must be taken not to spoil the gauge symmetry.
In our case, we need to describe QCD with the addition of
color-octet vector bosons. Following the prescription of
Ref.~\cite{ZR}, we introduce two massive vector fields, $\tilde
G^\mu$ and $\tilde\rho^\mu$, each transforming as an octet under
their respective $SU(3)_{G}$ and $SU(3)_{\rho}$ symmetry group,
with a mixing term which is fixed in such a way that the resulting
mass matrix has a zero eigenvalue. The corresponding eigenvector
is identified with the physical gluon field. The orthogonal
combination is identified with the physical color octet vector
resonance $\rho_{T8}$. The physical $SU(3)_{QCD}$ gauge symmetry
is the linear combination of the generators of $SU(3)_{G}$ and
$SU(3)_{\rho}$ corresponding to the massless gluon. The color
octet pseudoscalar $\eta_{T8}$ is introduced as a matter field
that transforms purely under gauged-$SU(3)_{\rho}$. In contrast,
quarks are introduced in the fundamental representation of the
gauged-$SU(3)_{G}$ symmetry. Hence, in the mass eigenbasis there
is a direct coupling of quarks to the physical color octet
resonance $\rho_{T8}$.

At this point the free parameters of the model can be taken as the
masses of the color octet vector and scalar particles,
$M_{\rho_{T8}}$ and $M_{\eta_{T8}}$, and a coupling constant
$g_\rho$ that controls the strong interaction decay of the
technirho into technietas, $\rho_{T_8} \rightarrow \eta_{T_8}
\eta_{T_8}$.

One extra parameter,  the PNGB decay constant  $F_Q$ is necessary to
describe the possible $\eta_{T8}$ decays.
The amplitude for $\eta_{T8} \rightarrow q \bar{q}$
decay is given by:
\begin{equation}
{\cal M} (\eta^a_{T8} \rightarrow q \bar{q}) = \frac{m_q}{F_Q} \bar{q}
\gamma_5 \frac{\lambda^a}{2} q
\end{equation}
where $\lambda^a$ are the Gell-Mann $SU(3)$ matrices.
The color octet technieta couples to gauge bosons through the
Adler-Bell-Jackiw anomaly with an amplitude given by \cite{anomaly}:
\begin{equation}
{\cal M}(\eta_{T8} \rightarrow B_1 B_2) = \frac{S_{\eta_{T8} B_1
    B_2}}{4 \sqrt{2} \pi^2 F_Q} \epsilon_{\mu \nu \alpha \beta}
 \varepsilon_1^\mu \varepsilon_2^\nu k_1^\alpha k_2^\beta
\end{equation}
where $\varepsilon_i$ and $k_i$ denote the polarization vector and
momentum of the vector boson $i$. For the cases we
are interested one has:
\begin{equation}
S_{\eta^a_{T8} g^b g^c} = g_s^2 d_{abc} N_{TC}
\end{equation}
and
\begin{equation}
S_{\eta^a_{T8} g^b \gamma} = \frac{g_s e}{3} \delta_{ab} N_{TC}
\end{equation}
where $N_{TC}$ is the number of technicolors, which we take as $N_{TC}=2$ for definiteness.

Hence, our model is completely determined by $M_{\rho_{T8}}$, $M_{\eta_{T8}}$,  $g_\rho$ and $F_Q$.
In the next section we perform a realistic simulation of the
possibility to detect a pair of color octet technietas produced in
the process $ pp \rightarrow \eta_{T8} \eta_{T8} \rightarrow \gamma g g g $ at the LHC.

\section{Simulation}

The model was implemented in CompHEP \cite{CompHEP}using LanHEP
\cite{LanHEP}. We used CompHEP for generating events for the
double technieta production. Subsequently, they were processed by
a FORTRAN code we wrote in order to generate the $\eta_{T8}$ decay
products. All the simulations were done for $M_{\eta_{T8}}=320$ GeV.
We restricted ourselves to consider only $M_{\eta_{T8}}$ below the top
quark threshold because above that point the technieta width is
strongly  depend on the existence of a topcolor interaction.

We considered two sources of Standard Model background: the production
of a photon
and three jets and the production of four jets
with the possibility that a jet can be misinterpreted as a photon. We
assume that the probability of such misidentification occurs is
about $10^{-3}$. The background was generated using Madgraph/MadEvent
with CTEQ5L as parton distribution function.
We considered gluons and all the (anti-)quarks of the first two generations
in the initial state and as sources of jets. All the possible tree
level partonic subprocesses were taken into account in the
calculation of $pp\rightarrow \gamma jjj$ and  $pp\rightarrow jjjj$.

In order to introduce some degree of realism in our calculations,
we took into account the smearing of the final momenta. We used
the following parametrization for the detector resolution:

\begin{eqnarray}
  \frac{\sigma(E)}{E}&=&\frac{0.20}{\sqrt{E(\gev)}}\;\;\; \mbox{ for
  photons}\\
\frac{\sigma(E)}{E}&=&\frac{0.80}{\sqrt{E(\gev)}}\;\;\; \mbox{ for jets}
\end{eqnarray}

In fact, it is expected that the resolution of ATLAS will be
better than what we used and in that sense our estimate must be
taken as conservative.

The reconstruction of the technieta  was done by studying the
photon-jet invariant mass ($M_{\gamma j}$) and the two jet
invariant mass ($M_{jj}$). We addressed the combinatorics problem
by taking the $M_{\gamma j}$ closer to $M_{\eta_{T_8}}$ as a
``technieta candidate''.

The background was reduced imposing the following set of kinematical
cuts:

\begin{eqnarray}
  P_{T\gamma}&>&80 \gev\\
  P_{T j}&>&80 \gev\\
  \left|M_{\gamma j}-M_{jj}\right|&<&0.15 M_{\eta}\\
 \left|M_{\gamma j}-M_{\eta_{T_8}}\right|&<&20 \gev.
\end{eqnarray}

\begin{figure}[hbt]
  \centering
  \includegraphics[scale=0.45]{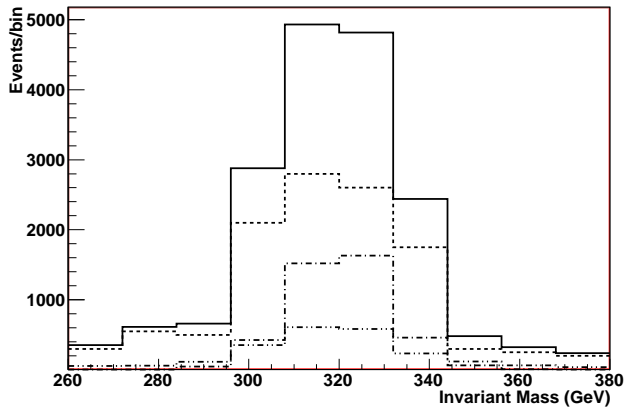}
  \caption{Reconstructed technieta invariant mass distribution The
    dashed line represents the direct background while the
    double-dotted-dashed line is the result obtained when a jet is
    misidentified as a photon. Our signal is shown by the
    dotted-dashed line.}
  \label{fig:thefigure}
\end{figure}

In figure \ref{fig:thefigure}, we show the reconstructed technieta
mass obtained by summing up the $M_{\gamma j}$ and $M_{jj}$
distributions after the cuts were applied. The dashed line
represents the direct background while the double-dotted-dashed
line is the result obtained when a jet is misidentified as a
photon. Our signal is shown by the dotted-dashed line.

Assuming an integrated luminosity of ${\cal L}=10$ fb$^{-1}$, we
expect to observe about 21400 events with 4200 of them
corresponding to our signal. That would correspond to a deviation
from the SM with a statistical significance of 32$\sigma$.

In this analysis, we have used some cuts that depend on the
technieta mass. This procedure may be uncomfortable from the
experimental point of view since the mass of the searched particle is
not known a priori and the whole possible range must be
scanned. Fortunately, in our case the mass interval is limited
because, due to QCD contributions, the technieta cannot be lighter
than 300 GeV and we do not expect that the channel considered in this
work to be useful for discovery if the technieta is heavier than 350 GeV.

However, it is possible to devise a search strategy which is
independent of the technieta mass. Consider the following set of
cuts:

\begin{eqnarray}
  P_{T\gamma}&>&80 \gev\\
  P_{T j}&>&80 \gev\\
  \left|M_{\gamma j}-M_{jj}\right|&<&10 \gev \\
 \left|\mbox{min}\left( \cos \theta_{\gamma j}\right)\right|&<&0.6
\end{eqnarray}
where $\theta_{\gamma j}$ is the angle formed by a photon and a
jet. The last cut comes from the fact that the technieta is a spin
0 particle and it decays isotropic in its rest system, while the
background tends to have peaks at $ \cos \theta_{\gamma j}=\pm 1$.

\begin{figure}[hbt]
  \centering
  \includegraphics[scale=0.45]{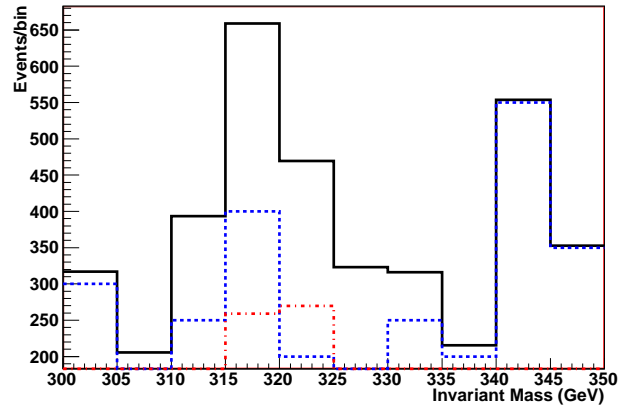}
  \caption{Reconstructed technieta invariant mass distribution using a
    set of cuts independent of the technieta mass. The
    dashed line represents the direct background
    (we neglect the small indirect background due to misidentification).
    Our signal is shown by the dotted-dashed line.}
  \label{fig:themasslessfigure}
\end{figure}

Figure \ref{fig:themasslessfigure} shows the invariant mass
distribution obtained with the new set of cuts. We expect to observe,
integrating over the whole mass range,
12600 background events and 1100 events coming from the technietas
with ${\cal L}=10$ fb$^{-1}$, corresponding to a $10 \sigma$ signal.

\section{Conclusions}
We studied the pair production and detection of color-octet
pseudoscalar bosons at the LHC, which could be present in models
of electroweak symmetry breaking induced by new strong
interactions. We restricted our analysis to $m_{\eta_{T8}}$ (the
color-octet pseudoscalar mass) below $2 m_t$, in which case the
decay into $t$-$\bar t$ is forbidden and consequently the
detection is more challenging. In such cases, the color-octet
pseudoscalar decays mainly into two gluons, induced by the
anomaly, while the direct decay into $b$-$\bar b$ has a fraction
less than 20\%. We perform simulations for the production and
decay of pseudoscalar pairs, including background. We did not look
for the dominant 4-jet mode, but for the suppressed 3-jet + photon
mode, which has a much lower background. We used two different
methods of analysis. In one method we assume $m_{\eta_{T8}}$ to be
known in our cuts, and obtain a number of events 32$\sigma$ above
the expected background, for an integrated luminosity of 10
fb$^{-1}$. In our second method, we did not include any value of
$m_{\eta_{T8}}$ in our cuts, but relied only on the invariant mass
reconstruction of the two decaying pseudoscalars, in which case
the number of events resulted in a statistical significance of
10$\sigma$ above the expected background. These results show that
the LHC has the potential to detect or exclude the existence of
such pseudoscalar colored bosons.


\section*{Acknowledgments}
The work of ARZ is partially supported by grant DID-UACH S-2006-28
and by Fondecyt grant 1070880; COD is partially supported by
Fondecyt, Chile, grant 1070227; RR is partially supported by a
CNPq, Brazil, research grant 309158/2006-0. CD and RR also
acknowledge the support of Fondecyt, Chile grant 7070098 for
international cooperation. This work is also supported by a CNPq
PROSUL program 490157/2006-8.

\end{document}